# Standing spin wave excitation in Bi:YIG films via temperature induced anisotropy changes and magnetoacoustic coupling

Steffen Peer Zeuschner[1,2], Xi-Guang Wang[3,4], Marwan Deb[1], Elena Popova[5], Gregory Malinowski[6], Michel Hehn[6], Niels Keller[5], Jamal Berakdar[4] and Matias Bargheer[1,2,*]

[1] Institut für Physik und Astronomie, Universität Potsdam, Karl-Liebknecht-Str. 24-25, 14476 Potsdam, Germany

[2] Helmholtz Zentrum Berlin für Materialien und Energie GmbH, Albert-Einstein-Str. 15, 12489 Berlin, Germany

[3] School of Physics and Electronics, Central South University, Changsha 410083, China

[4] Institut für Physik, Martin-Luther Universität, Halle-Wittenberg, Halle/Saale D-06120, Germany

[5] Institut de Physique de Rennes (IPR), CNRS UMR6251 Univ Rennes, 35000 Rennes, France

[6] Institut Jean Lamour (IJL), CNRS UMR 7198, Université de Lorraine, 54506 Vandœuvre-lès-Nancy, France

*e-mail : bargheer@uni-potsdam.de

Based on micromagnetic simulations and experimental observations of the magnetization and lattice dynamics following the direct optical excitation of the magnetic insulator Bi:YIG or indirect excitation via an optically opaque Pt/Cu double layer, we disentangle the dynamical effects of magnetic anisotropy and magnetoelastic coupling. The strain and temperature of the lattice are quantified via modeling ultrafast x-ray diffraction data. Measurements of the time-resolved magneto-optical Kerr effect agree well with the magnetization dynamics simulated according to the excitation via two mechanisms: The magneto-acoustic coupling to the experimentally verified strain dynamics and the ultrafast temperature-induced transient change in the magnetic anisotropy. The numerical modeling proves that for direct excitation both mechanisms drive the fundamental mode with opposite phase. The relative ratio of standing spin-wave amplitudes of higher order modes indicates that both mechanisms are substantially active.

**Introduction:**

Information processing at GHz and THz frequencies entails a combination of photonic and magnetic channels, and hence the optical manipulation of collective spin excitations at high frequencies.(1, 2) Magnetic insulators are particularly suitable as the optical absorption is much lower than for metals and the damping of spin waves may be engineered to be very low. An example are Yttrium Iron Garnet (YIG) thin films which have become a test bed for the optical control of magnetization dynamics.(3, 4) Bismuth doping (resulting in Bi:YIG) enhances substantially the magneto-optical contrast while affecting only slightly the elastic parameters and the magnetic damping.(4–6) Recent experiments confirmed that the magnetization dynamics, in particular standing spin waves in thin films can be directly excited simultaneously by one- and two photon absorption in Bi:YIG.(7, 8) Under strong excitation conditions, the frequency of the fundamental mode is persistently changed by 20%.(9) Also indirect excitation of Bi:YIG via an adjacent opaque metal heterostructure can trigger standing spin waves.(10) Experimental evidence suggests that coherent or incoherent phonons entering Bi:YIG (propagating strain waves or diffusive phonon heat) are responsible for triggering the spin wave motion. Ultrafast x-ray diffraction experiments have been used to quantify the spatio-temporal profile of strain waves and phonon temperature changes under both excitation conditions.(11) While the two dominant mechanisms for exciting spin waves, namely the magnetoacoustic coupling and the temperature induced anisotropy changes, are surely active under the experimental conditions, it has never been quantitatively tested, how they individually affect the magnetization dynamics and what is their relative importance.

Here we analyze and contrast our experimental and simulation results to unveil quantitatively the relationship between the observed strain waves and/or the heat input and the standing spin waves in nanometric magnetic insulator films, which are directly or indirectly excited via ultrashort laser-pulses. Ultrafast x-ray diffraction (UXRD) data calibrate the spatio-temporal strain and the temperature profile in a Bi:YIG film after i) direct optical excitation of the insulating Bi:YIG, and ii) indirect excitation via an optically opaque Pt/Cu double layer. These experimentally verified spatio-temporal strain and heat profiles are adopted as source terms in the micromagnetic model based on the Landau-Lifshitz-Gilbert (LLG) equation. The simulation results are compared to time-resolved magneto-optical Kerr effect (tr-MOKE) measurements of the spin-wave excitations in the same thin films. The relative ratio of the two mechanisms can be experimentally verified by the amplitude ratio of the excited higher order modes to the fundamental mode, because this mode is excited with opposite phase by the two mechanisms, which therefore partially cancel out. The strain is composed of propagating strain waves and quasi static thermal strain following the diffusion of heat. It couples to the magnetization via a linear magnetoacoustic coupling. In addition, the temperature induces local changes of the anisotropy constant. Consequently, both strain and temperature modulate the effective magnetic field on ultrafast timescales triggering standing spin waves.

The paper is structured as follows: First we discuss the micromagnetic model and an analytic solution of the LLG equation in section (A). In section (B) we show the experimental tr-MOKE signal for direct and indirect excitation. In section (C) we derive the spatiotemporal strain and temperature maps, $T(z,t)$ and $\eta(z,t)$ from the UXRD data. With these experimentally verified input data, the tr-MOKE signal is modeled numerically in section (D) according to the LLG equation. The model is compared to the experimental tr-MOKE data from section (B) for direct and indirect excitation of the magnetic layer. In both cases we separately model the influence of the two inputs, the spatio-temporal temperature, and the strain.

### A) Micromagnetic model and analytical considerations

Before discussing the experimental observations and the results of full numerical modeling, we analyze a linearized analytical solution of the Landau-Lifshitz-Gilbert (LLG) equation(12) without damping and without dynamic dipole-dipole interaction. Here, only transversal spin dynamics that do not change the length of the magnetization vector is of interest. Hence, the magnetization field $\vec{M}$ can be described by a unit field vector $\vec{m}$ with $\vec{M} = M_s \vec{m}$ and $M_s$ is the saturation magnetization. The time evolution of $\vec{m}$ is governed by $\partial_t \vec{m} = -\gamma \, \vec{m} \times \vec{H}_{eff}$, where the effective field $\vec{H}_{eff}$ follows from the functional derivative of the magnetization free-energy density with respect to $\vec{m}$ and is detailed below. For a small amplitude precession $\vec{m} = \vec{m}_0 + \delta\vec{m}_{sw}$ around the equilibrium $\vec{m}_0$ with $|\vec{m}_0| \gg |\delta\vec{m}_{SW}|$ one obtains to a linear order in $|\delta\vec{m}_{SW}|$ an analytical solution of the LLG:(13) In our case, the external field is perpendicular to the film and its field strength $H_z$ can be set by an electromagnet. The effective field $\vec{H}_{eff}$(14) acting on the magnetization vector is a sum of the contributions(15) listed in Table 1. (the spatial and the time dependence of $\vec{m}$ are suppressed for brevity)

$$\vec{H}_{eff}(z,t) = H_z \vec{e}_z + \left[H_d + H_u(z,t) + H_{me}(z,t) + H_c(z,t)\left(m_x^2 + m_y^2\right)\right]m_z \vec{e}_z + H_c(z,t)\left[(m_y^2 + m_z^2)m_x \vec{e}_x + (m_x^2 + m_z^2)m_y \vec{e}_y\right] + D_{ex}\frac{\partial^2 \vec{m}}{\partial z^2} \; . \tag{1}$$

The magnetoelastic coupling $H_{me}(z,t) = -\frac{2b_1 \eta_{zz}(z,t)}{\mu_0 M_s}$ depends on time and space via the spatio-temporal strain $\eta_{zz}(z,t)$.(10) The uniaxial and cubic anisotropy constants $K_x(T)$ are temperature dependent, and therefore the anisotropy fields $H_u(z,t) = -\frac{2}{\mu_0 M_s}K_x(T(z,t))$ depend on time and space via the spatiotemporal temperature profile $T(z,t)$.(16) The external field $H_z$ along the $\vec{e}_z$ is partially compensated by the large uniaxial ($H_u$) anisotropy field and the demagnetization field $H_d = -N_z M_s$, where $N_z = 1$ is the demagnetization factor of the thin film geometry. The compensation is proportional to the z-component of the unit magnetization vector $m_z$. Since $m_z$ grows with $H_z$ as the magnetization tilts out of plane, the effective field would be zero without the cubic anisotropy as long as $H_z < H_s$ (see Fig. 1(a) for illustration). Thus, the small cubic ($H_c$) essentially determines the direction of the magnetization at low external fields. When $H_z$ reaches the saturation field (critical field) $H_s$, above which the magnetization is aligned along $\vec{e}_z$, the effective field grows linearly with the external field (Kittel mode). This field $H_s$ is determined by the minimum of the free energy, including cubic anisotropy.

| | constants at RT | Effective field | Effective Field (in mT/$\mu_0$) |
|---|---|---|---|
| Cubic Anisotropy | $K_c = 400$ J/m$^3$ | $H_c = \dfrac{2K_c(T(z,t))}{\mu_0 M_s}$ | $H_c = 5.9$ |
| Uniaxial Anisotropy | $K_u = 5750$ J/m$^3$ | $H_u = -\dfrac{2K_u(T(z,t))}{\mu_0 M_s}$ | $H_u = -85$ |
| Demagnetization field / saturation magnetization | $M_s = 170$ mT | $H_d = -N_z M_s$ | $H_d = -170$ |
| Saturation Field | | $H_s = -(H_u + H_d)$ | $H_s = 260$ |
| Magnetoelastic field(17) | $b_1 = 3.5 \cdot 10^5$ J/m$^3$ | $H_{me} = -\dfrac{2b_1 \eta_{zz}(z,t)}{\mu_0 M_s}$ | $H_{me} = 62$ per 1% strain $H_{me} = 8$ per $\Delta T = 100K$ |
| Change of uniaxial anisotropy with T(18) | | | $\Delta H_u = 31$ per $\Delta T = 100K$ |
| Exchange constant | $A_{ex} = 3.5 \cdot 10^{-12}$ J/m | $D_{ex} k_z^2 = \dfrac{2A_{ex}}{\mu_0 M_s} k_z^2$ | |
| Gyromagnetic ratio | $\dfrac{\gamma}{2\pi} = 28$ GHz/T | | |

**Table1: Magnetic properties of the Bi:YIG film** according to the simultaneous fit of the lower branches for the modes n = 0, … 4. The second column collects formulas converting magnetic constants to effective fields which enter the LLG equation. The third column compares the values of the time-dependent effective fields $\Delta H_u$ (per strain) and $H_{me}$ (per strain and per $\Delta T$) to the magnitude of the static effective fields. Note that the largest contributions $H_d m_z$ and $H_u m_z$ are canceled by the external filed $H_z$ in the perpendicular geometry chosen here.

Before excitation, i.e., in the stationary state with $\dfrac{\partial m}{\partial t} = 0$, the state $\vec{m}_0 = (0,0,1)$ is stable, if $H_z > -H_d - H_u + H_c = H_s$ (upper branch or Kittel mode – see Fig. 1(c). In wavevector space, the exchange term $D_{ex} \dfrac{\partial^2 \vec{m}}{\partial z^2}$ transforms to $D_{ex} k_z^2$ for a mode with a wavevector $k_z$. Submitting $\vec{m} = \vec{m}_0 + \delta \vec{m}_{sw}$ into the LLG equation, we obtain to linear order in $|\delta \vec{m}_{sw}|$ the frequency on the external field.

$$\omega_u(H_z) = \gamma(D_{ex} k_z^2 + H_z + H_d + H_u - H_c). \qquad (2)$$

If $H_z < H_s$ (lower branch) the stable magnetization is at an angle $\theta_M$ with respect to the surface normal $\vec{m}_0 = (\sin\theta_M \cos\varphi_M, \sin\theta_M \sin\varphi_M, \cos\theta_M)$, and $\varphi_M$ describes the in-plane rotation of $\vec{m}$. Here, $H_z + \left(H_d + H_u + \dfrac{H_c}{2}\right)\cos\theta_M - \dfrac{3H_c}{2}\cos^3\theta_M = 0$ and $\varphi_M = \dfrac{\pi}{4}$. In this case the frequency is

$$\omega_l(H_z) = \gamma \sqrt{(D_{ex} k_z^2 + H_c \sin^2\theta_M)\left(D_{ex} k_z^2 - \left(H_d + H_u - 4H_c \cos^2\theta_M + \dfrac{1}{2} H_c \sin^2\theta_M\right)\sin^2\theta_M\right)}$$
(3)

For the fundamental mode ($k_z = 0$), the frequency $\omega_l(H_z)$ of the lower branch would be strictly zero without cubic anisotropy (see first factor of the radicand), if the external field $H_{ext}$ has no in-plane components ($H_{ext} = H_z$). $H_z$ would then be perfectly cancelled out by $(H_d + H_u)\cos\theta_M$. Thus, the cubic anisotropy is crucial for obtaining a finite frequency. Eq. (3) exhibits only an implicit dependence of the frequency on the external field which determines the direction of the magnetization parametrized by $\theta_M$. See Fig. 1(a) for an illustration of the complete cancellation of the out-of plane components of the effective field at the saturation field. For $H_z \geq H_s$, the cubic anisotropy contributions to the effective field disappears [Fig. 1(b)] according to Eq. (1), because $m_x$ and $m_y$ vanish.

Thus, there are two branches of the spin wave frequency [Fig. 1(c)] as a function of the external field, which both depend on the wavevector component along the external field. For the thin film, the out-of plane wavevector component is quantized: $k_z = \dfrac{n\pi}{d}$. This spatial quantization of the spin waves essentially enters their frequency via the exchange term $D_{ex} k_z^2$ in Eq. (2) and (3). In Fig. 1(c) we fit equation Eq. (3) to the lower branch of the experimental data from Deb et al.(7) We obtain a convincing agreement for the lower branches for $n = 0, \dots, 4$ under the assumption of the parameters given in Table 1, which agree reasonably well with the literature. For the upper branch, the magnetization points perfectly out of plane. Hence, the spin wave amplitude is exceedingly small for a perfect alignment of the external field, because the excitation is forbidden by symmetry. This and a tiny experimental in-plane field may explain the deviations.

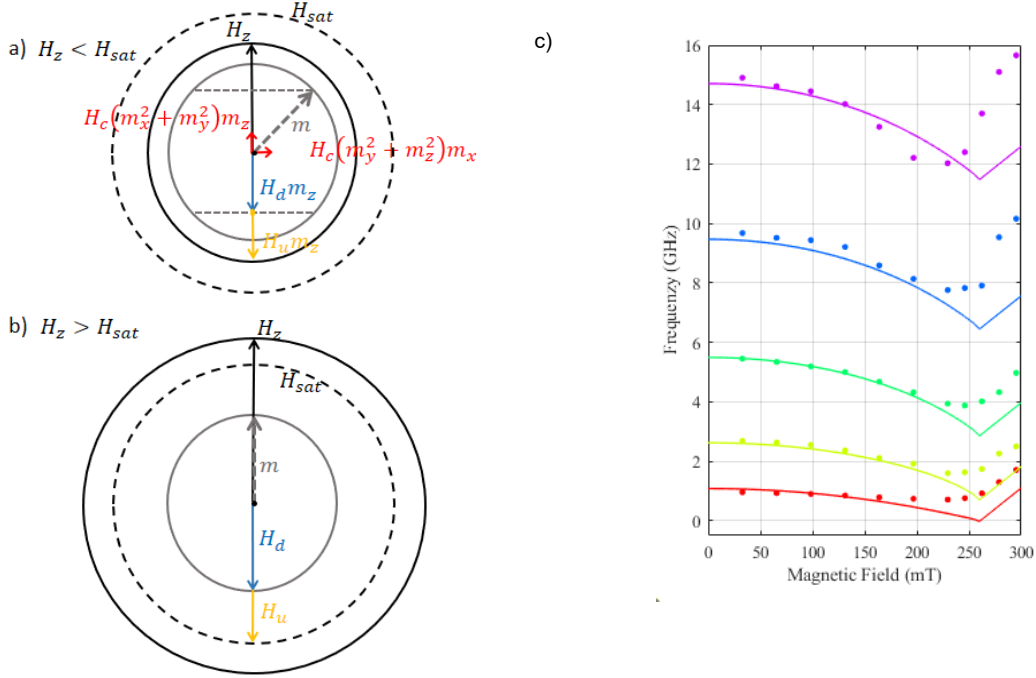

Fig. 1: a) Vector diagram of the contributions to the effective field for $H_z < H_s$. The cubic anisotropy $H_c$ determines the effective field and hence the stable direction of the magnetization vector (see the contributions to the effective field in Eq. (1), because $H_d m_z + H_u m_z$ nearly cancel $H_z$. b) For $H_z \geq H_s$ the magnetization points along $\vec{e_z}$. At the saturation field ($H_z = H_s$), all effective field contributions cancel to zero. c) Spin wave frequencies according to Eq. (2) and (3) for the first five standing spin wave modes (n=0,…,4). All branches fit well simultaneously to the tr-MOKE data from Deb et al., i.e. they fit the quadratic dependence on the mode number.(7)

**B) Experimental spin wave spectrum for direct and indirect excitation**

In this paper we study a pair of $Bi_1Y_2Fe_5O_{12}$ (Bi:YIG) single crystalline thin films with a thickness of 135 nm grown by pulsed laser deposition on a gadolinium gallium garnet (GGG) (100) substrate.(19, 20) One of the two films is covered by a Pt (5 nm)/Cu (100 nm) bilayer on top of the Bi:YIG film using direct current magnetron sputtering. During this procedure the thickness of the Bi:YIG is reduced by few nanometers. The top Pt layer absorbs nearly all photons from the optical excitation pulses and the 100 nm Cu layer ensures that no photons reach the Bi:YIG layer. However, the metallic bilayer is known to efficiently conduct the absorbed photon energy by efficient heat transport via hot electrons to the magnetic film. For metallic ferromagnets, the hot electron pulse would directly funnel energy into its electronic system.(21) However, the electrons are stopped at the interface to the ferromagnetic insulator Bi:YIG, and energy is transported in the ferromagnet via coherent and incoherent phonons.

In the following we compare the magnetization dynamics triggered in the Bi:YIG film by energy directly deposited in the bulk of the uncovered Bi:YIG film to the scenario, where the energy is transported to the Bi:YIG interface via the Cu film, excluding direct optical excitation. Fig. 1(c) we reproduce the measurement of the standing spin wave modes upon laser-excitation using tr-MOKE.(7) The standard setup has been described previously.(22) In the left column of Fig. 2 we show the direct absorption of 800 nm pulses in the Bi:YIG film (by one and two photon absorption) and the right column of Fig. 2 shows the results for indirect excitation, where the 800 nm light pulses are fully absorbed in the Pt/Cu layers with 5nm/100nm thickness. Figs. 2(a) and 2(b) show schematics for direct and indirect excitation of the Bi:YIG film by pump pulses with a wavelength of the 800 nm. The Kerr rotation of a 400 nm probe pulse detects the magnetization dynamics Fig. 2(c) and 2(d) show exemplary tr-MOKE traces under the two excitation conditions. The Fourier-transforms of these spectra are shown in 2(e) and 2(g), which clearly show the same higher order spin wave modes at slightly different frequencies, because of the different thickness of the two Bi:YIG films. We link these exemplary spectra to the full experimental determination of the dependence of the frequency spectrum on the external field in f), which shows the quadratic dependence of the frequency on the mode number. The solid lines show fits through the data in f) using Eq. (3), which contains a linear and quadratic contribution in $n = \frac{k_z d}{\pi}$ because we fit the lower branch [Eq. (3)].

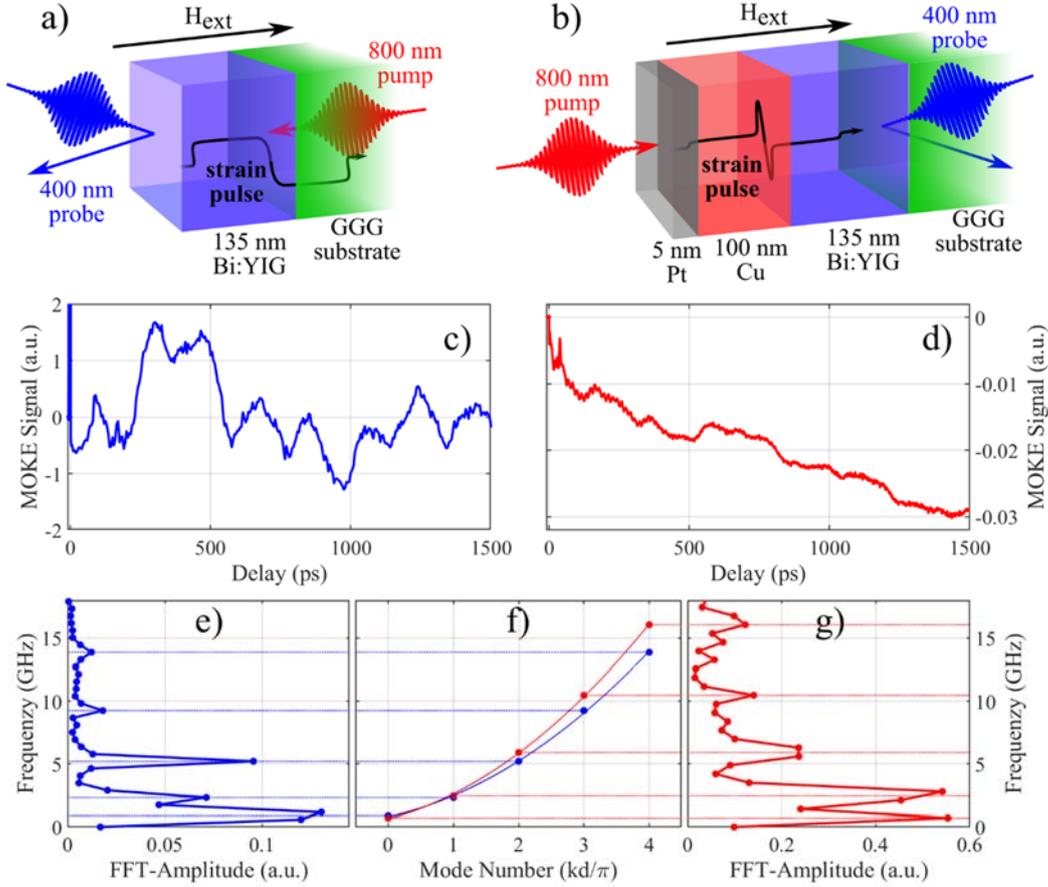

**Fig. 2) Measurement of standing spin waves by MOKE** a) Schematic of the direct excitation geometry, where 800 nm pump pulses at 10 mJ/cm² incident fluence are absorbed in Bi:YIG. b) Schematic of the indirect excitation geometry. c) Tr-MOKE for an external field of $\mu_0 H_z = 98$ mT perpendicular to the sample for direct excitation. d) Time resolved MOKE at $\mu_0 H_z = 88$ mT for indirect excitation. Fourier-transforms of the spectra exhibiting five resonance frequencies shown for the direct e) and indirect g) excitation. f) Measured spin wave frequencies for direct (blue) and indirect (red) excitation extracted from the maxima in panels e) and g). The quadratic dependence of the mode number is given by the exchange energy [cf. Eq. (3)], and is slightly steeper for the indirect excitation, where the Bi:YIG film is thinner and the external field is smaller (higher frequencies in the lower branch of Fig. 1(c).

### C) Experimental quantification of the strain and temperature by UXRD

The standing spin waves reported in Fig. 2 are excited by laser-induced coherent and incoherent phonons, meaning by strain waves and phonon heating. The transient strain $\eta_{zz}(z,t)$ including the quasi-static strain due to thermal expansion couples to the spins via the magnetoacoustic coupling term $H_{me}(z,t)$. The temperature rise $T(z,t)$ induces changes of the magnetocrystalline anisotropy ($H_u(z,t)$ and $H_c(z,t)$), triggering spin precession, too. Fig. 3 summarizes the characterization of strain waves and heat transport by UXRD as reported by Zeuschner et al.(11) We probe the strain in the Bi:YIG layer by the shift of the Bragg peak via reciprocal space slicing(23) at our laser-driven x-ray plasma source.(24) In analogy to Fig. 2, the left column of Fig. 3 shows direct excitation and the right column shows indirect excitation which is identical to the conditions for the tr-MOKE experiments in Fig.2, i.e. in the indirect excitation all light is absorbed in the metal layers. The x-ray probe pulses penetrate the entire heterostructure and we can determine the average transient strain $\bar{\eta} = \frac{\Delta c}{c}$ by following the angular shift of the Bragg peak [dots in Fig. 3 (a) and 3(d)]. The average transient strain $\bar{\eta}(t)$ is the direct observable in our experiments. On the other hand, for time-delays beyond 100 ps the propagating strain waves have left the Bi:YIG layer and the remaining quasi-static strain $\bar{\eta}$ is a direct measure of the average temperature change in the Bi:YIG layer. A thorough modeling of the transient strain (solid lines) gives access to the spatio-temporal temperature gradient $T(z,t)$ and transient strain $\eta(z,t)$.(11, 21) We calculate the energy deposited in the Bi:YIG layer and in the heterostructure from the incident fluence. The diffusive two temperature model is required to correctly account for the non-equilibrium heat transport in the Pt/Cu bilayer.(11, 21) We use the thermo-elastic parameters given in Table 2.

The spatio-temporal strain [Fig. 3(b) and 3(e)] and heat [Fig. 3(c) and 3(f)] maps serve as input parameters for the simulation of the magnetization dynamics. Since we recorded the MOKE and UXRD data in different setups, with different focusing conditions and laser pulse parameters, we scale the strain in the UXRD data according to the absorbed fluence. For the indirect excitation the absorption of light in the metals and the heat flow are nearly linear in the excitation fluence. The direct excitation was shown to have linear and quadratic contributions.(7, 11) We scaled the absorbed fluence to match the absorption of the light pulses observed in the MOKE setup. The modeling confirms that the observed strain and heat do not significantly depend on the relative weight of the linear and quadratic absorption contributions.

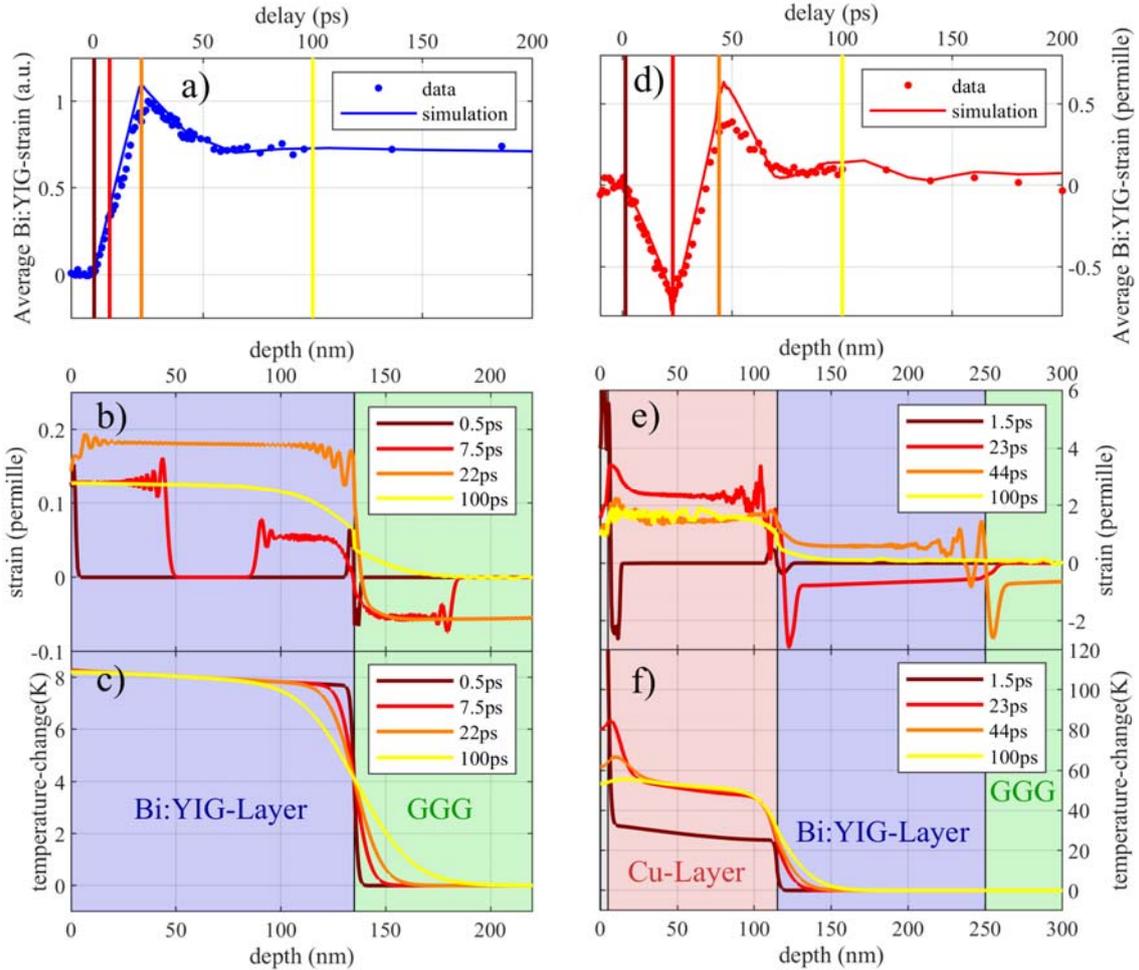

**Fig. 3 Calibration of strain and heat by UXRD**

a) Transient strain in Bi:YIG following direct excitation at 800 nm extracted from Zeuschner et al.(11) and scaled to the fluence 10 mJ/cm² used in the tr-MOKE experiment. The blue line is a fit according to the model. The spatial dependence of the calculated strain b) and temperature c) is shown for the selected times indicated by the color code in a). The right panels d),e) and f) show the same for indirect excitation, where the data are reproduced from Deb et al.(8) Note the larger temperature rise (f) in the metal films and the strong gradient in the Bi:YIG layer. The average temperature in Bi:YIG is smaller for the indirect excitation, but the standing spin wave modes are efficiently excited by the gradient. The simulations of the magnetization dynamics in Fig. 4 use only the spatiotemporal strain and temperature in the blue shaded areas, which symbolize the Bi:YIG layers.

In the following we comment briefly on the main observations for strain and heat. Fig. 3(b) illustrates, that upon direct excitation the Bi:YIG layer expands via two counter-propagating strain fronts starting at the surface and the interface. At 22 ps both strain fronts have reached the opposite side of the film, the maximum average strain is reached and the Bi:YIG layer is nearly homogeneously expanded. In the next 22 ps, the expansive strain front that started at the GGG interface is converted to a contraction wave at the surface. Then, the acoustic dynamics cease as both strain fronts have propagated into the substrate. In the Bi:YIG layer, the quasi-static expansion given by the thermal load in the film remains and very slowly relaxes via thermal transport to the substrate. Fig. 3(c) illustrates, that already on short timescales the heat flow yields a temperature gradient at the interface, which is already considerable after about 100 ps.

For the indirect excitation, Fig 3(d) illustrates that the strain in Bi:YIG film is essentially driven by a bipolar strain pulse, which results from an expansion of the Cu layer. Unlike one may have expected, the short bipolar strain pulse with high amplitude generated in Pt does not considerably alter the average strain in Bi:YIG. The simulation clearly shows that the very localized bipolar pulse from Pt, which has three times larger an amplitude than the compression launched by Cu, travels through Bi:YIG (see e.g. the snapshot at 23 ps in Fig. 3(e). However, the Bi:YIG strain is essentially determined by the Cu transducer. We emphasize that electrons transport energy to the Cu/Bi:YIG interface quasi instantaneously (see 1.5 ps snapshot in Fig. 3(f) and hence start the relevant acoustic compression of Bi:YIG immediately, as indicated by the experiment [Fig. 3(d)]. After the dominant passage of the bipolar strain in the first 75 ps, only a small expansion of Bi:YIG is observed due to the slow phononic heat transport into the Bi:YIG layer, which is quantified in Fig. 3(f). Again, at 100 ps, a considerable temperature gradient is built, however in this case, it is located at the interface to Cu and not towards the substrate.

| Property at 300K | GGG | Bi:YIG |
| --- | --- | --- |
| Sound velocity (nm/ps) | **6.34** (25) | **6.3** (26) |
| Density (g/cm³) | **7.085** (27) | **5.9** (26) |
| Linear thermal expansion ($10^{-5}$/K) | **1.65** (28) | 1 |
| Heat Capacity (J/kg/K) | **381.2** (29) | 560 |
| Poisson-Correction $(1 + 2C_{12}/C_{11})$(30) | | 1.8 |
| Thermal conductivity (W/m K) | 7.05 | 7.4 |

**Table. 2 Thermoelastic parameters of Bi$_1$Y$_2$IG and the substrate GGG:** The bold parameters can be found for Bi:YIG in the given references. The other parameters are assumed reported for YIG and assumed to be similar for    Bi:YIG. Parameters for Pt and Cu are identical to the ones used in the publication from Pudell et al.(21)

### D) Numerical micromagnetic modeling that uses strain and temperature from UXRD

Fig. 4) summarizes the results of the micromagnetic simulations according to the model introduced in section A). The full LLG equation is solved numerically based on the parameters given in Table 1 and the explicit spatiotemporal dependence of the strain $\eta_{zz}(z,t)$ and the temperature $T(z,t)$, which were derived in section C) by modeling the lattice strain calibrated by the UXRD data. In analogy to Fig. 2 and 3 the left column of Fig. 4 collects simulation results for the direct excitation of Bi:YIG and the right column for indirect excitation. The only difference in the simulations stems from the different UXRD calibrated input values for strain $\eta_{zz}(z,t)$ and temperature $T(z,t)$.

The upper row displays results, where the magnetoacoustic coupling constant $b_1$ is set to zero and only the temperature driven anisotropy change $\Delta H_u(z,t) \propto \Delta H_c(z,t) \propto T(z,t)$ are at work (we use the linearized dependence $\Delta H_u(T)$ from Table 1 for parametrization.(5, 18) In the second row only the magnetoacoustic coupling launches spin waves, because we set the anisotropy to be temperature independent, i.e. constant. While all calculated spectra show the expected frequency spectrum, we clearly see the excitation of the fundamental mode upon direct excitation to be much stronger than in the experiment. The calculated magnetization dynamics (insets in Figs 4(a) and 4(c) display $m_z(z,t)$) show that the fundamental mode is triggered with opposite phase for the two mechanisms. In order to achieve at least a qualitative agreement with tr-MOKE data, both mechanisms must be active. Both mechanisms individually trigger a much larger fundamental mode amplitude compared to the higher order modes. Because the precise values for magnetoacoustic coupling coefficients and anisotropy changes of our samples may differ from the literature values,(5, 18) we scale the relative magnitude of the magnetoacoustic coupling with respect to the anisotropy mechanism $H_{me}/\Delta H_u = c\, H_{me}^{lit}/\Delta H_u^{lit}$ by a coefficient $c$.

For the literature value ($c = 1$) the fundamental mode has much too large amplitude. We obtain reasonable agreement with the experimentally observed spectra in Fig. 2(c) for direct excitation, by using $c \approx 3$. The modeling of the indirect excitation depends much less on the parameter $c$. The experiments rely on the same Bi:YIG film and are only different in terms of excitation. The main difference between the direct and indirect excitation is that for the indirect excitation the strain essentially travels through the Bi:YIG film quickly. The temperature change in Bi:YIG induced by the heat flow from the metal to the film has a very strong spatial gradient, which imposes a strong spatial gradient not only on $\Delta H_u$, but also on the effective field change, $H_{me}^{qs}$ due to the quasistatic change of the strain by thermal heating.

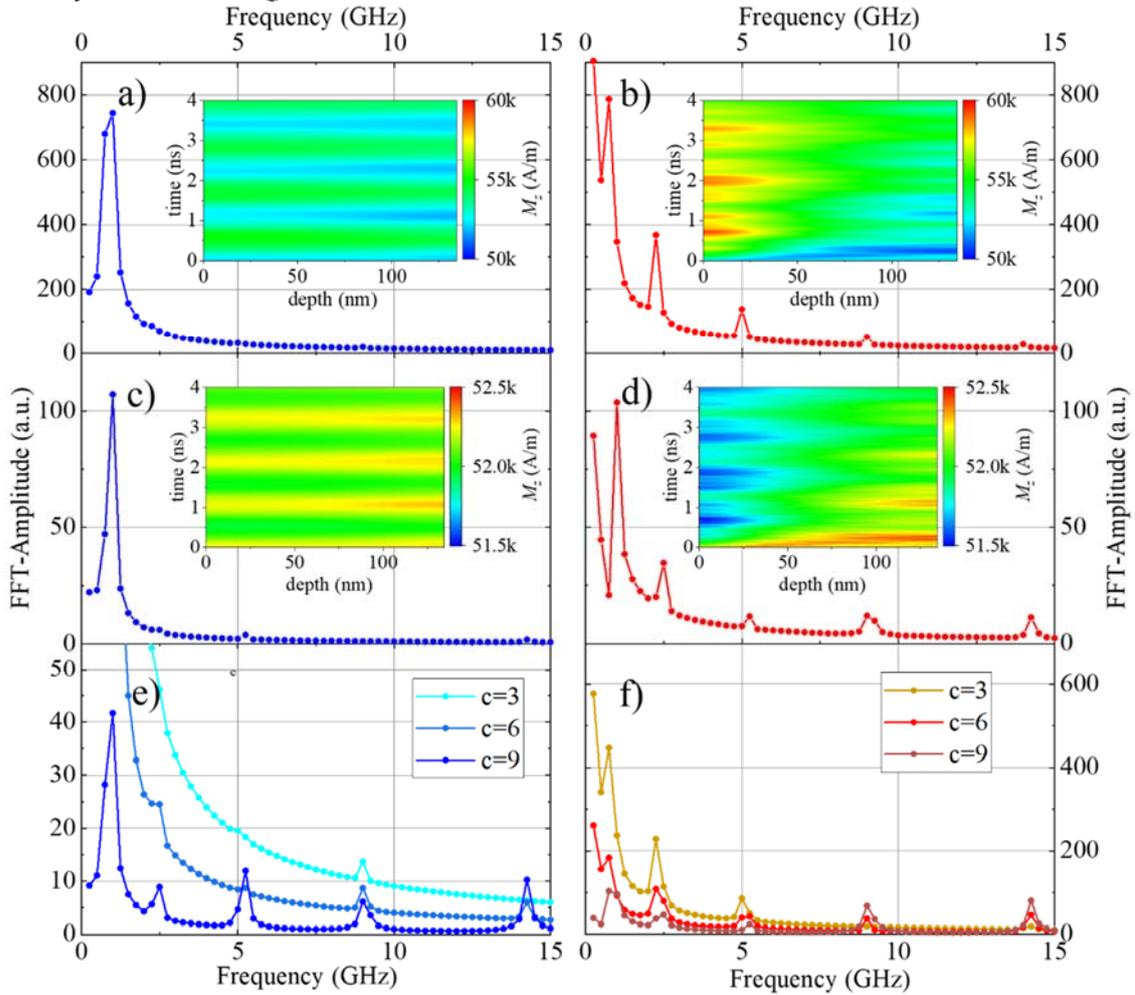

**Fig. 4 Numerical micromagnetic modeling based on the input of $\eta_{zz}(z,t)$ and $T(z,t)$ from experimental UXRD data**
Spin-wave spectra resulting from the Fourier-analysis of the calculated $m_z(z,t)$ shown in the respective inset. The first row shows the result, were the magnetoacoustic coupling is zero ($H_{me} = 0$) and only the temperature induced anisotropy change is active ($H_u = -\frac{2K_u(T(z,t))}{\mu_0 M_s}$). a) direct excitation b) indirect excitation. For the second row only the magnetoacoustic term $H_{me} = -\frac{2b_1 \eta_{zz}(z,t)}{\mu_0 M_s}$ is on, whereas the anisotropy is constant ($H_u = -\frac{2K_u}{\mu_0 M_s}$). c) direct excitation d) indirect excitation. From the insets, which show $m_z(z,t)$ we see that the two mechanisms drive the fundamental mode with opposite phase (see panels a) and c)). In the last row both mechanisms are active and the parameter c quantifies the relative strength of the magnetoacoustic coupling. e) for the direct excitation and $c \approx 3$ the fundamental mode is considerably lowered, and the calculated spectrum resembles the experimental observation in Fig. 2(f). For the indirect excitation (f) the suppression of the fundamental mode does not exceed the suppression of the higher modes. For the choice of the same $c \approx 3$ the calculated spectra are again in reasonable agreement with experiment [Fig. 2(g)].

**Discussion**

The modeling of UXRD data strictly determines the connection of the transient strain profile $\eta(z,t)$ and the transient temperature $T(z,t)$. For direct excitation, the temperature change leads to a nearly homogeneous quasi-static $\eta_{QS} = \alpha \Delta T$ determined by the temperature rise, where $\alpha = 10^{-5} 1/K$ is the thermal expansion coefficient. The effective magneto-elastic coupling field $H_{me}^{qs} = 8$ mT due to the quasi-static strain $\eta_{QS}(100K)$ is about four times smaller than $\Delta H_u = 31$ mT for $\Delta T = 100$ K. This temperature-induced quasi static strain necessarily enters the effective field, and hence the temperature changes $\Delta T$ act on the effective field in two ways. The anisotropy route is about four times more efficient, however. We note that in addition to the quasi-static strain, the ultrafast optical excitation generates strain waves which may dominate the spin wave excitation if the acoustic wave timescale is resonant with the spinwave. For the indirect excitation the strain wave is launched in the Pt/Cu structure, and hence the contribution of the magneto-elastic coupling can be relatively larger, because the related temperature change is high in the metal but less pronounced in Bi:YIG. Hence the temperature induced changes in Bi:YIG contribute less, the thicker the Cu film is. The magnetoacoustic strain wave has two bipolar components, which are given by the Pt and Cu thickness [Fig. 3(e)]. Both transients enter the LLG equation via the effective field given in Eq. (1). The numerical modeling has essentially only two tuning parameters: The change of the anisotropy constant with temperature $\Delta H_u(T)$ and the magnetoelastic coupling $H_{me}(\eta)$. The change of the magnetocrystalline anisotropy due to magnetostriction(31) is negligible in the case of YIG and therefore Bi:YIG which leads to independent excitation routes of SSWs in Bi:YIG: Thermally induced anisotropy change via ultrafast heating on the one hand and picosecond strain waves and thermal expansion paired with inverse magnetostriction on the other hand.

In our modeling of the spin-wave spectrum we tune the relative magnitude of the magnetoacoustic coupling coefficient. We have tested that the magnetic response to both excitations is still in the linear regime. We find that the experimental spin wave spectrum for direct excitation of Bi:YIG in Fig. 2(c) can only be reproduced for a relatively strong magnetoacoustic coupling. The key is that the zero order mode (ferromagnetic resonance (FMR) mode) is excited with opposite phase for the two mechanisms. If we neglect either the magneto-acoustic coupling [Fig. 4(a)] or the temperature-induced anisotropy change [Fig. 4 (c)] the higher order modes have a much too small amplitude compared to the FMR mode. For an appropriately increased magnitude of the magneto-elastic coupling ($c \approx 3$), or alternatively a reduced $\Delta H_u(T)$, we arrive at a simulated spin-wave spectrum that approximately reproduces the measured one.

**Conclusion**

We have set up a micromagnetic model to quantitatively test the relative importance of strain waves and temperature changes for the excitation of spin waves. We find that the two mechanisms counteract each other for the FMR mode, but may have a different phase relationship for higher order modes. We can quantitatively test our model by calibrating the input parameters of transient strain and transient phonon temperature via UXRD and comparing the simulated spin wave spectrum with experimental result obtained in MOKE experiments under the same excitation conditions. We have verified the model for both cases, the indirect excitation of Bi:YIG via a metal bilayer on the one hand, as well as by one and two-photon absorption on the other hand. We believe that this analysis of the two coupled routes of generating magnetization dynamics via coherent and incoherent phonons is widely applicable for other ferromagnetic, ferrimagnetic and antiferromagnetic materials and may help to design new devices.

**Acknowledgement:**

This work was funded by the Deutsche Forschungsgemeinschaft (DFG, German Research Foundation) – Project-ID 328545488 – TRR 227, projects A10 and B06. X.-G.W. thanks the National Natural Science Foundation of China (Nos. 12174452, 12074437, 11704415), and the Natural Science Foundation of Hunan Province of China (No. 2021JJ30784) for support. M.D. acknowledges the Alexander von Humboldt Foundation for financial support.